\documentclass[useAMS,usenatbib,usegraphicx]{mn2e}

\usepackage{epsf}
\usepackage{amssymb,amsmath}

\title[Testing the gas mass profile of galaxy clusters with DDR]
{Testing the gas mass density profile of galaxy clusters with
distance duality relation }

\author[Cao et al.]
{Shuo Cao$^{1}$, Marek Biesiada$^{1,2}$, Xiaogang Zheng$^{1}$, and Zong-Hong Zhu$^{1}$\\
$^1$ Department of Astronomy, Beijing Normal University, 100875, Beijing, China; \emph{zhuzh@bnu.edu.cn}\\
$^2$ Department of Astrophysics and Cosmology, Institute of Physics,
University of Silesia, Uniwersytecka 4, 40-007 Katowice, Poland }

\begin{document}

\date{\today}

\voffset- .5in

\pagerange{\pageref{firstpage}--\pageref{lastpage}} \pubyear{}

\maketitle

\label{firstpage}

\begin{abstract}
In this paper, assuming the validity of distance duality relation,
$\eta=D_L(z)(1+z)^{-2}/D_A(z)=1$, where $D_A(z)$ and $D_L(z)$ are
the angular and the luminosity distance respectively, we explore two
kinds of gas mass density profiles of clusters: the isothermal
$\beta$ model and the non-isothermal double-$\beta$ model. In our
analysis, performed on 38 massive galaxy clusters observed by
\textit{Chandra} (within the redshift range of $0.14<z<0.89$), we
use two types of cluster gas mass fraction data corresponding to
different mass density profiles fitted to the X-ray data. Using two
general parameterizations of $\eta(z)$ (phenomenologically allowing
for distance duality violation), we find that the non-isothermal
double-$\beta$ model agrees better with the distance duality
relation, while the isothermal $\beta$ model tends to be marginally
incompatible with the Etherington theorem at 68.3\% CL. However,
current accuracy of the data does not allow to distinguish between
the two models for the gas-density distribution at a significant
level.
\end{abstract}


\section{Introduction}\label{sec:introduction}

The distance duality relation (DDR thereafter): $ D_{
L}/D_A(1+z)^{-2}= 1$, which relates the luminosity distance $D_L(z)$
to the angular diameter distance $D_A(z)$ at a given redshift $z$,
is known to hold in any metric theory of gravity. Hence, it has
extensively been applied in modern observational
cosmology\citep{Schneider92,Bassett04,CMS07,Zhu08b,Mantz13,Holanda10,Cao11a}.
Until now, different astronomical observations have been used to
check the validity of this reciprocity relation using the following
expression
\begin{equation}
  \frac{D_{\scriptstyle L}}{D_{\scriptstyle A}}{(1+z)}^{-2}=
  \eta(z),
  \label{rec}
\end{equation}
where $\eta(z)$ could be a function of the redshift $z$ (of course
DDR corresponds to $\eta(z)=const.=1$). For instance, \citet{Uzan04}
used the $D_A$ estimated from Sunyaev-Zeldovich effect (SZE) and
X-ray surface brightness of 18 galaxy clusters \citep{Reese02}, and
found DDR consistent with the observations at 1$\sigma$ CL. The
validity of the distance duality relation was further confirmed by
\citet{bem06}, with a larger sample of angular diameter distances
from 38 galaxy clusters \citep{Boname06}. More recently, instead of
testing the reciprocity relation itself, \citet{Holanda10} discussed
the possibility of using DDR (i.e. assuming that it is true) to test
the geometrical shape of galaxy clusters. They used two angular
diameter distance $D_A$ measurements based on two different cluster
geometries: ellipsoidal $\beta$ model underlying \citet{Filippis05}
data and spherical $\beta$ model assumed by \citep{Boname06}. They
found that ellipsoidal geometry was more consistent with the DDR,
and concluded that it was a better model to describe galaxy
clusters.

In this paper, we propose a new method to derive observed
$\eta_{obs}(z)$ parameter from the cluster gas mass fraction $f_{\rm
gas} = M_{\rm gas}/M_{\rm tot}$ inferred from X-ray data, and we use
the assumption that DDR should be valid (i.e. $\eta(z) = 1$) to
discuss appropriateness of two cluster gas mass density profiles
assumed. More specifically, we use two data sets of cluster gas mass
fraction derived from \textit{Chandra} X-ray data \citep{LaRoque06}
under two different assumptions about the gas mass density profiles:
isothermal $\beta$ model and non-isothermal double-$\beta$ model.

In our analysis we consider two particular parameterizations of
phenomenological $\eta(z)$ dependence: I. $\eta= \eta_{0} +
\eta_{P1} z$; II. $\eta (z) = \eta_{0} + \eta_{P2} z/(1+z)$. The
first expression is a linear parametrization equivalent to the first
order Taylor expansion in redshift. The second one is inspired by
the commonly used CPL parametrization for dark energy equation of
state \citep{linder04} and is equivalent to the first order
expansion in the scale factor $a(t)$ which is the only gravitational
degree of freedom in Freedman - Robertson - Walker cosmology. These
two parametrizations have been extensively used to investigate the
properties of dark energy in the literature
\citep{Cao12,Cao14,Cao15}. Assuming that the Etherington theorem is
valid (which is quite a reasonable assumption), the best-fit value
obtained with a given data set should be $\eta_0=1$ and $\eta_P=0$.
Our results indicate that, the non-isothermal double-$\beta$ model
tends to be more compatible with the reciprocity relation than the
isothermal $\beta$ model. This kind of result is an interesting
example of how general principles (like DDR) could be used to assess
the validity of assumptions concerning local physical conditions.

This paper is organized as follows. In Section~\ref{sec:Sample} we
present two samples of gas mass fraction data from 38 X-ray luminous
galaxy clusters and their corresponding mass density profiles.
Statistical method and constraint results on $\eta(z)$ parameters
are shown in Section~\ref{sec:analysis}. Finally, we summarize our
main conclusions and make a discussion in
Section~\ref{sec:Conclusions}.

\section{Galaxy Cluster Samples }\label{sec:Sample}

The cluster gas mass fraction is defined as a ratio of the X-ray
emitting gas mass to the total mass of a cluster, i.e., $f_{\rm gas}
= M_{\rm gas}/M_{\rm tot}$. Gas mass $M_{\rm gas}$ is derived from
the X-ray surface brightness while the total mass $M_{\rm tot}$ can
be obtained by assuming that the gas is in hydrostatic equilibrium
with the cluster NFW 
potential. In order to perform appropriate calculations, one needs
the gas mass density profile. Therefore, in our analysis aimed at
constraining free parameters ($\eta_{0}$, $\eta_{P}$) in general
expression for $\eta(z)$, we will use the gas mass fractions
obtained with different gas mass density profiles.

In order to calculate $\eta_{obs}$ from the data (according to
Eq.(\ref{eq:eta2}) shown below) we use two sets of gas mass fraction
$f_{\rm gas}$ both derived from a sample of 38 luminous X-ray
clusters with temperatures $T_{\rm gas}> 5$ keV, observed by
\textit{Chandra} X-ray Observatory \citep{LaRoque06} located at
redshifts from $z=0.14$ to $z=0.89$. In fact, LaRoque et al. (2006)
presented results of their analysis using both X-ray only data (for
isothermal $\beta$-model) and a combination of \textit{Chandra}
X-ray data and BIMA/OVRO interferometric radio SZE data for the
double - $\beta$ model. However, according to \citet{Holanda12}, the
gas mass fraction measurement via SZE depends only on the angular
diameter distance, so it is insensitive to the validity of the DDR.
Therefore, only the high resolution \textit{Chandra} X-ray data are
considered in our analysis. The above mentioned two sets of $f_{\rm
gas}$ were obtained within the most commonly used isothermal $\beta$
model and under assumption of the non-isothermal double - $\beta$
model, respectively.

\begin{figure}
\begin{center}
\includegraphics[width=1.0\hsize]{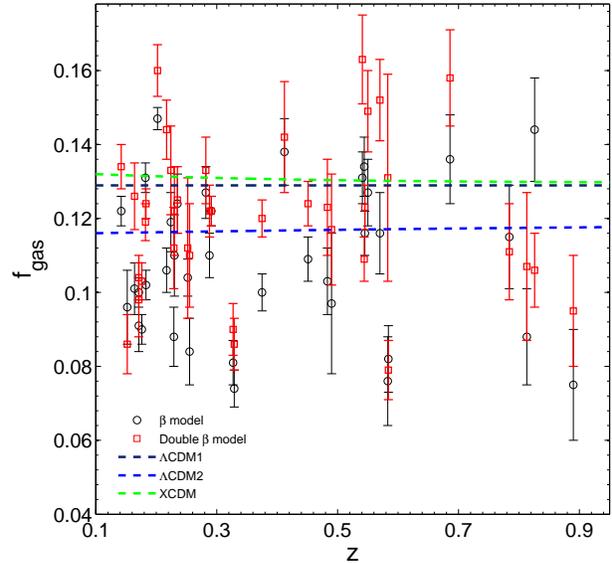}
\end{center}
\caption{ The cluster gas mass fraction data derived from a sample
of 38 luminous X-ray clusters observed by \textit{Chandra} X-ray
Observatory \citep{LaRoque06}. Black circles and red squares with
the error bars correspond to the results obtained under the
assumption of isothermal $\beta$ model and non-isothermal
double-$\beta$ model, respectively. The predictions of $f_{gas}(z)$
under the assumption of DDR are also overplotted, for the three
reference cosmologies (see Table~\ref{tab:cosmology}) considered in
this paper. \label{data}}
\end{figure}

\begin{figure}
\begin{center}
\includegraphics[width=1.0\hsize]{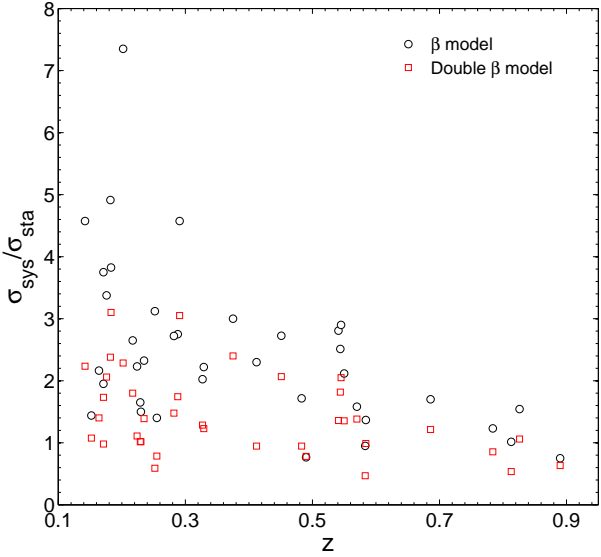}
\end{center}
\caption{ The ratio between systematic and statistical uncertainties
for the cluster gas mass fraction data. Black circles and red
squares with the error bars correspond to the results obtained under
the assumption of isothermal $\beta$ model and non-isothermal
double-$\beta$ model, respectively. \label{ratio}}
\end{figure}

Fig.~\ref{data} displays these data and their observational
statistical uncertainties. We will also consider possible systematic
uncertainties influencing the derived gas mass fraction following
the discussion in \citep{Boname06} concerning their method of
calculating angular diameter distances to clusters. The effect of
these systematical uncertainties on the gas mass fractions is
summarized in Table~3 of \citet{LaRoque06}. Note that the systematic
difference between $f_{\rm gas}$ derived from the isothermal $\beta$
model and the double - $\beta$ model still exists. From a purely
statistical point of view, non-parametric Wilcoxon signed rank test
reveals that difference between these two datasets is significant at
the level of $p=0.0001$. More specifically, the assumption of
isothermality can potentially affect the gas mass fraction
measurements by about $(-5\%)$ through its effects on both gas mass
and total mass. Therefore, an appropriate systematic uncertainty
will be assigned to the gas mass fractions derived from the
isothermal $\beta$ model. In order to show the major sources of
uncertainty in the present analysis, we display the ratio between
systematic and statistical uncertainties for these data in
Fig.~\ref{ratio}.

In the frequently used isothermal $\beta$ model, the 3-dimensional
electron number density $n_e$ can be written as
\citep{Cavaliere76,Cavaliere78,Grego01,Reese02,Ettori04}
\begin{equation}
n_e(r) = n_{e0} \left ( 1 + \frac{r^2}{r_c^2} \right )^{-3\beta/2},
\label{eq:single_beta}
\end{equation}
where $n_{e0}$ is the central electron number density, $r$ is the
radius from the center of the cluster, $r_c$ is the core radius of
the intracluster medium (ICM), and $\beta$ is a power law index.
Although it has been widely used as a useful assumption in galaxy
cluster studies, the isothermal $\beta$ model still encounters some
problems with describing the central emission excess seen in some
clusters \citep{Mohr99,LaRoque06} and the X-ray surface brightness
at $r>(1 - 1.5)\;r_{2500}$, where $r_{2500}$ is the radius at which
the mean enclosed mass density is equal to $2500 \rho_{crit}$. Such
features, non-compatible with the isothermal $\beta$ model have been
observed in archival ROSAT PSPC data and numerical simulations
\citep{Mohr99,Borgani04}.

Therefore, in order to overcome the limitations of the simplest
isothermal $\beta$ model we have also considered a more
sophisticated cluster plasma model, the non-isothermal
double-$\beta$ model. More specifically, its density profile is
assumed to be of the following form:
\begin{equation}
n_e(r) = n_{e0} \left[ f \left ( 1 + \frac{r^2}{\,r_{c_1}^2}\right)
^{-3\beta/2} + (1-f)\left (1 +
\frac{r^2}{\,r_{c_2}^2}\right)^{-3\beta/2}\right]
\label{eq:double_beta}
\end{equation}
where $r_{c_1}$ is the core radius representing the narrow, peaked
central density component contributing to the central density
$n_{e0}$ by a factor $0\leq f \leq 1$,  and $r_{c_2}$ is the other
core radius describing the broad, shallow outer density profile.

\begin{table*}
\caption{\label{tab:cosmology} Parameters of three cosmologies
considered: standard concordance $\Lambda$CDM model ($\Lambda$CDM1),
$\Lambda$CDM best fitted to Planck data ($\Lambda$CDM1), XCDM model
best fitted to Planck+WMAP9 data. }
\begin{center}
\begin{tabular}{l|c|c|lllllllllllllll}\hline\hline
Cosmology        &  Cosmological parameters \\
\hline $\Lambda$CDM1  & $\Omega_m=0.30$, $H_0=70.0 km s^{-1}Mpc^{-1}$  \\
\hline $\Lambda$CDM2   &  $\Omega_m=0.315$, $H_0=67.3 km s^{-1}Mpc^{-1}$  \\
\hline XCDM            & $\Omega_m=0.294$, $w=-1.05$, $H_0 = 70.4 km s^{-1}Mpc^{-1}$  \\

\hline \hline

\end{tabular}
\end{center}
\end{table*}

Let us now describe how we extracted the observed DDR parameter
$\eta_{obs}$ from the cluster gas mass fraction data. Massive galaxy
cluster are the largest known bound systems in the Universe and are
expected to provide a unique information of the matter content of
the Universe. The baryonic-to-total mass ratio of clusters should
closely match the ratio of $\Omega_b/\Omega_m$, where $\Omega_b$ and
$\Omega_m$ are the present dimensionless density parameter of the
baryonic matter and dust matter, respectively. Since the
reconstructed $f_{gas}$ depends on the angular diameter distances at
different redshifts, $f_{gas} \propto D_{A}^{1.5}(z)$, a number of
studies have used it to probe the acceleration of the Universe and
thus the properties of dark energy
\citep{Allen04,Allen08,Ettori09,Cao14}.

\begin{table*}
\caption{\label{tab:result} Summary of the results for
$\eta(z)=\eta_0+\eta_{P1}z$ and $\eta(z)=\eta_0+\eta_{P2}z/(1+z)$,
respectively, at 1$\sigma$ confidence level for the $\beta$ model
and double-$\beta$ model. The statistical and systematic
uncertainties in $\eta$ are shown separately. Two samples including
the $n=38$ full sample and the $n=29$ sub-sample are used (see text
for definitions). Source of cosmological priors are also given in
brackets. }

\begin{center}
{\scriptsize
\begin{tabular}{l|c|c|lllllllllllllll}\hline\hline
$\beta$ model: Parameterizations (Sample/Cosmology)        & $\eta_0$   &   $\eta_{P1}$ \\
\hline $\eta(z)=\eta_0+\eta_{P1}z$ (Full sample/$\Lambda$CDM1) & $\eta_0=1.068\pm0.099(stat)\pm0.034(sys)$    & $\eta_{P1}=-0.084\pm0.058(stat)\pm 0.129(sys)$  \\
\hline $\eta(z)=\eta_0+\eta_{P2}z/(1+z)$ (Full sample/$\Lambda$CDM1) & $\eta_0=1.080\pm0.104(stat)\pm0.051(sys)$    & $\eta_{P2}=-0.148\pm0.117(stat)\pm 0.276(sys)$  \\
\hline $\eta(z)=\eta_0+\eta_{P1}z$ (Sub-sample/$\Lambda$CDM1) & $\eta_0=1.108\pm0.115(stat)\pm0.033(sys)$    & $\eta_{P1}=-0.144\pm0.075(stat)\pm0.312(sys)$  \\

\hline $\eta(z)=\eta_0+\eta_{P1}z$ (Sub-sample/$\Lambda$CDM2) & $\eta_0=1.028\pm0.104(stat)\pm0.037(sys)$    & $\eta_{P1}=-0.127\pm0.072(stat)\pm 0.127(sys)$  \\
\hline $\eta(z)=\eta_0+\eta_{P2}z/(1+z)$ (Sub-sample/$\Lambda$CDM2) & $\eta_0=1.060\pm0.112(stat)\pm0.047(sys)$    & $\eta_{P2}=-0.276\pm0.134(stat)\pm 0.282(sys)$  \\

\hline $\eta(z)=\eta_0+\eta_{P1}z$ (Sub-sample/XCDM) & $\eta_0=1.242\pm0.127(stat)\pm0.041(sys)$    & $\eta_{P1}=-0.265\pm0.087(stat)\pm 0.148(sys)$  \\
\hline $\eta(z)=\eta_0+\eta_{P2}z/(1+z)$ (Sub-sample/XCDM) & $\eta_0=1.285\pm0.135(stat)\pm0.052(sys)$    & $\eta_{P2}=-0.529\pm0.166(stat)\pm 0.305(sys)$  \\

\hline \hline

Double-$\beta$ model: Parameterizations (Sample/Cosmology)        & $\eta_0$   &   $\eta_{P2}$ \\
\hline $\eta(z)=\eta_0+\eta_{P1}z$ (Full sample/$\Lambda$CDM1) & $\eta_0=0.991\pm0.102(stat)\pm0.015(sys)$    & $\eta_{P1}=-0.037\pm0.060(stat)\pm 0.087(sys)$  \\
\hline $\eta(z)=\eta_0+\eta_{P2}z/(1+z)$ (Full sample/$\Lambda$CDM1) & $\eta_0=1.002\pm0.104(stat)\pm0.027(sys)$    & $\eta_{P2}=-0.092\pm0.114(stat)\pm 0.178(sys)$  \\
\hline $\eta(z)=\eta_0+\eta_{P1}z$ (Sub-sample/$\Lambda$CDM1) & $\eta_0=0.981\pm0.104(stat)\pm0.017(sys)$    & $\eta_{P1}=-0.023\pm0.069(stat)\pm 0.089(sys)$  \\
\hline $\eta(z)=\eta_0+\eta_{P2}z/(1+z)$ (Sub-sample/$\Lambda$CDM1) & $\eta_0=0.996\pm0.108(stat)\pm0.030(sys)$    & $\eta_{P2}=-0.078\pm0.132(stat)\pm 0.184(sys)$  \\

\hline $\eta(z)=\eta_0+\eta_{P1}z$ (Sub-sample/$\Lambda$CDM2) & $\eta_0=0.915\pm0.095(stat)\pm0.018)(sys)$    & $\eta_{P1}=-0.015\pm0.064(stat)\pm 0.082(sys)$  \\
\hline $\eta(z)=\eta_0+\eta_{P2}z/(1+z)$ (Sub-sample/$\Lambda$CDM2) & $\eta_0=0.924\pm0.102(stat)\pm0.024(sys)$    & $\eta_{P2}=-0.049\pm0.121(stat)\pm 0.182(sys)$  \\

\hline $\eta(z)=\eta_0+\eta_{P1}z$ (Sub-sample/XCDM) & $\eta_0=1.101\pm0.117(stat)\pm0.024(sys)$    & $\eta_{P1}=-0.139\pm0.078(stat)\pm 0.08(sys)$  \\
\hline $\eta(z)=\eta_0+\eta_{P2}z/(1+z)$ (Sub-sample/XCDM) & $\eta_0=1.123\pm0.121(stat)\pm0.027(sys)$    & $\eta_{P2}=-0.257\pm0.143(stat)\pm 0.196(sys)$  \\

\hline \hline

\end{tabular}}
\end{center}
\end{table*}

Our starting point is the following general expression for the gas
mass fraction (see e.g. \citet{Allen08, Holanda10}):
\begin{eqnarray}
\label{eq:fraction}
 f_{\rm gas}(z)&=& KA\Upsilon
\left(\frac{\Omega_{\rm b}}{\Omega_{\rm m}}\right)
\left(\frac{D_L^{\rm ref}(z)}{D_L(z)}\right) \left(\frac{D_A^{\rm
ref}(z)}{D_A(z)}\right)^{1/2} \nonumber \\
   &=& KA\Upsilon \left(\frac{\Omega_{\rm b}}{\Omega_{\rm m}}\right) \eta_{obs}^{-3/2}
\left(\frac{D_A^{\rm ref}(z)}{D_A(z)}\right)^{3/2}
\end{eqnarray}
\noindent Here, $\eta_{obs}$ is the value of the DDR parameter
implied by observations. $K$ is a calibration constant
characterizing the systematic uncertainty on the overall
normalization \citep{Mantz13}.
We use 
$K = 0.94 \pm 0.09$ based on recent results from weak lensing mass
measurements \citep{Applegate13,Allen13}. $A$ is an angular
correction factor quantifying the shift of the angular scale
$\theta^{\rm }_{2500}$ of the $r_{2500}$ radius from $\theta^{\rm
ref}_{2500}$ obtained in the reference cosmology assumed to be
$\Lambda$CDM \citep{Mantz13}. So, the $A$ factor reads:
\begin{equation}
A =  \left(\frac{\theta^{\rm ref}_{2500}}{\theta^{\rm
}_{2500}}\right)^{\varepsilon} \sim \left(\frac{H(z) D_{\rm
A}(z)}{H^{\rm ref}(z) D_{\rm A}^{\rm
ref}(z)}\right)^{\varepsilon}\,. \label{eq:angcosm}
\end{equation}
\noindent where the exponent $\varepsilon$  assessed for the
0.8--1.2\,$\theta_{2500}$ shell \citep{Mantz13,Allen13} is
$\varepsilon=0.442 \pm 0.035$. $\Upsilon$ is the gas depletion
parameter related to thermodynamic history of X-ray emitting gas in
the course of cluster formation. More specifically, this factor is
modeled as $\Upsilon_{2500}=\Upsilon_{\rm 0}(1+\alpha_{\rm
\Upsilon}z)$. We adopt uniform priors, $\Upsilon_0 = 0.845 \pm
0.042$ and $\alpha_{\rm \Upsilon} = 0.00\pm0.05$ in our paper. This
is motivated by the recent hydrodynamic simulations of the hottest
clusters within 0.8--1.2\,$r_{\rm 2500}$ shells
\citep{Battaglia12,Planelles13}. $\Omega_b$ is fixed at the best-fit
value $\Omega_b h^2 = 0.02205 \pm 0.00028$ suggested by Planck
observations \citep{Ade14}. $D_{\rm A}(z)$ denotes the true angular
diameter distance and $D_{\rm A}^{\rm ref}(z)$ is the angular
diameter distance calculated in the reference cosmological model,
which is a flat $\Lambda$CDM model with $\Omega_m=0.30$, $h=0.70$
(standard concordance model).

By combing Eq.(\ref{eq:fraction}) and Eq.(\ref{eq:angcosm}), we can
obtain the observed value of $\eta_{obs}$ as
\begin{equation}
\label{eq:eta2} \eta_{obs}^{3/2}=  K\Upsilon_{\rm 0}(1+\alpha_{\rm
\Upsilon}z)\frac{\Omega_{\rm b}}{\Omega_{\rm m}}f_{\rm gas}^{-1}
\left(\frac{H(z)}{H^{\rm ref}(z)}\right)^{\varepsilon} \left(
\frac{D_A^{\rm ref}(z)}{D_A(z)}\right)^{3/2-\varepsilon}.
\end{equation}
It is apparent that in our analysis, $\eta(z)$ function includes two
effects: possible DDR violation and our uncertainty with respect to
the true cosmological model. Since our main goal is to test cluster
gas mass density profiles based on the validity of distance duality
relation, $D_A(z)$ will be calculated in three ways. Firstly, by
assuming the standard model with $\Omega_m=0.30$ and $H_0=70.0 km
s^{-1}Mpc^{-1}$, which is equivalent to the reference cosmology in
cluster studies and very similar to the WMAP9 constraints
\citep{Komatsu10}. Secondly, by taking the best-fit matter density
parameter and Hubble constant given by Planck Collaboration:
$\Omega_m=0.315$ and $H_0=67.3 km s^{-1}Mpc^{-1}$ \citep{Ade14}, in
the framework of $\Lambda$CDM model. And thirdly, by considering the
XCDM model, i.e. the one in which the equation of state $p = w \rho$
for dark energy has constant $w$ parameter. More precisely, we take
the best fitted parameters from Planck+WMAP9 data: $\Omega_m=0.294$,
$w=-1.05$, and $H_0 = 70.4 km s^{-1}Mpc^{-1}$ \citep{Cai14}. Summary
of three cosmologies adopted here can be found in
Table~\ref{tab:cosmology}. For comparison, the predictions of
$f_{gas}(z)$ within these three cosmologies under the assumption of
DDR and using the central values of four nuisance parameters are
also shown in Fig~\ref{data}.

\section{Analysis and Results}\label{sec:analysis}

As we already mentioned, $\eta(z)$ function captures
phenomenologically our uncertainty about not only the DDR but also
about true cosmological model and possibly other systematics.
Therefore it is reasonable to treat it as a function of redshift and
start with two quite natural parameterizations:
\begin{equation} \label{eq:1}
\left\{ \begin{aligned}
\eta(z)& = \eta_0+\eta_{P1}z, \\
\eta(z)& = \eta_0+\eta_{P2}z/(1+z).
\end{aligned} \right.
\end{equation}
where $\eta_0$ and $\eta_P$ parameters quantify the shift from the
expected standard result ($\eta_0=1$, $\eta_P=0$).

Using routines available within CosmoMC package \citep{Lewis02}, we
preformed Monte Carlo simulations of the posterior likelihood ${\cal
L} \sim \exp{(- \chi^2 / 2)}$, where
\begin{equation}
\label{chi2} \chi^{2} = \sum \frac{{\left[\eta(z)- \eta_{obs}(z)
\right] }^{2}}{\sigma^2_{\eta_{obs}} },
\end{equation}
$\eta_{obs}(z)$ was calculated from the $f_{gas}$ data via
Eq.(\ref{eq:eta2}) and $\sigma_{\eta_{obs}}$ was calculated
according to the standard law of uncertainty propagation. Besides
the statistical errors of X-ray observations, we have also
considered systematic errors concerning instrument calibration $\pm
6\%$, X-ray background $+2\%$, hydrostatic equilibrium $-10\%$, and
isothermal assumption $-5\%$ \citep{LaRoque06}. Combined in
quadrature, they result in a typical relative error of 10\% for the
$f_{gas}$ measurements with isothermal $\beta$ model and 15\% for
the $f_{gas}$ measurements with the non-isothermal double-$\beta$
model. We would stress however, that according to recent numerical
simulations and comparisons between X-ray and lensing masses
\citep{Lau09,Landry13,Giles15}, the hydrostatic mass underestimates
the true mass especially at large radii (which causes the gas mass
fraction to be overestimated), and the estimate of systematic
uncertainties due to hydrostatic equilibrium can be larger than
10\%. This effect still needs to be investigated with more available
data. In our analysis, based on the results of
\citet{Lau09,Landry13}, a systematic uncertainty of +10\% on the
total mass $M_{\rm tot}$ is assessed for all clusters, which
corresponds to a systematic error of -10\% on $f_{gas}$.

We performed a MCMC analysis and marginalized over the nuisance
parameters ($K$, $\Upsilon_0$, $\alpha_{\rm \Upsilon}$,
$\varepsilon$) by multiplying the probability distribution functions
and then integrating \citep{Ganga1997}. When $D_A(z)$ is calculated
within the fiducial model $(\Omega_{\rm m}=1-\Omega_{\Lambda}, H_0)
= (0.30, 70 km s^{-1}Mpc^{-1})$ consistent with WMAP9 observations,
we obtain the results shown in Table~\ref{tab:result}. Obviously
both statistical and systematic uncertainties should be included in
the analysis. Therefore they have been displayed explicitly in
Table~\ref{tab:result}. In the case of the first parametrization
$\eta(z)= \eta_0+\eta_{P1}z$, the best-fit $\eta$ parameters are
$\eta_0=1.068\pm0.133$, $\eta_{P1}=-0.084\pm 0.187$ (sta+sys) for
the isothermal $\beta$ model and $\eta_0=0.991\pm0.117$,
$\eta_{P1}=-0.037\pm 0.147$ (sta+sys) for the non-isothermal
double-$\beta$ model. One can see that gas mass fractions obtained
from the non-isothermal double beta model, are in better agreement
with the reciprocity relation ($\eta_{0} = 1$, $\eta_{P1} = 0$).
However, from the statistical point of view this  preference is
marginal.

\begin{figure}
\begin{center}
\includegraphics[width=1.0\hsize]{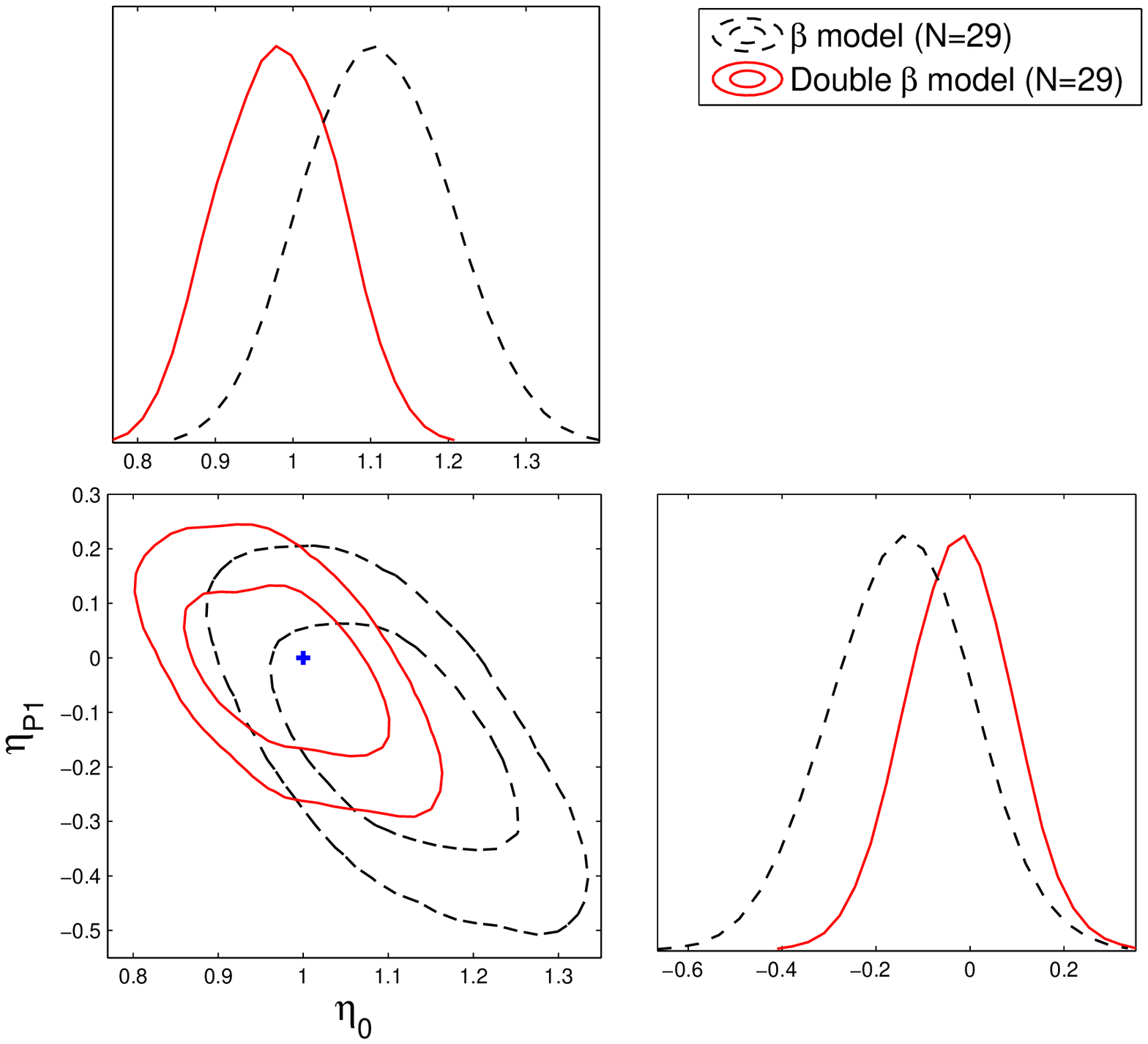}
\includegraphics[width=1.0\hsize]{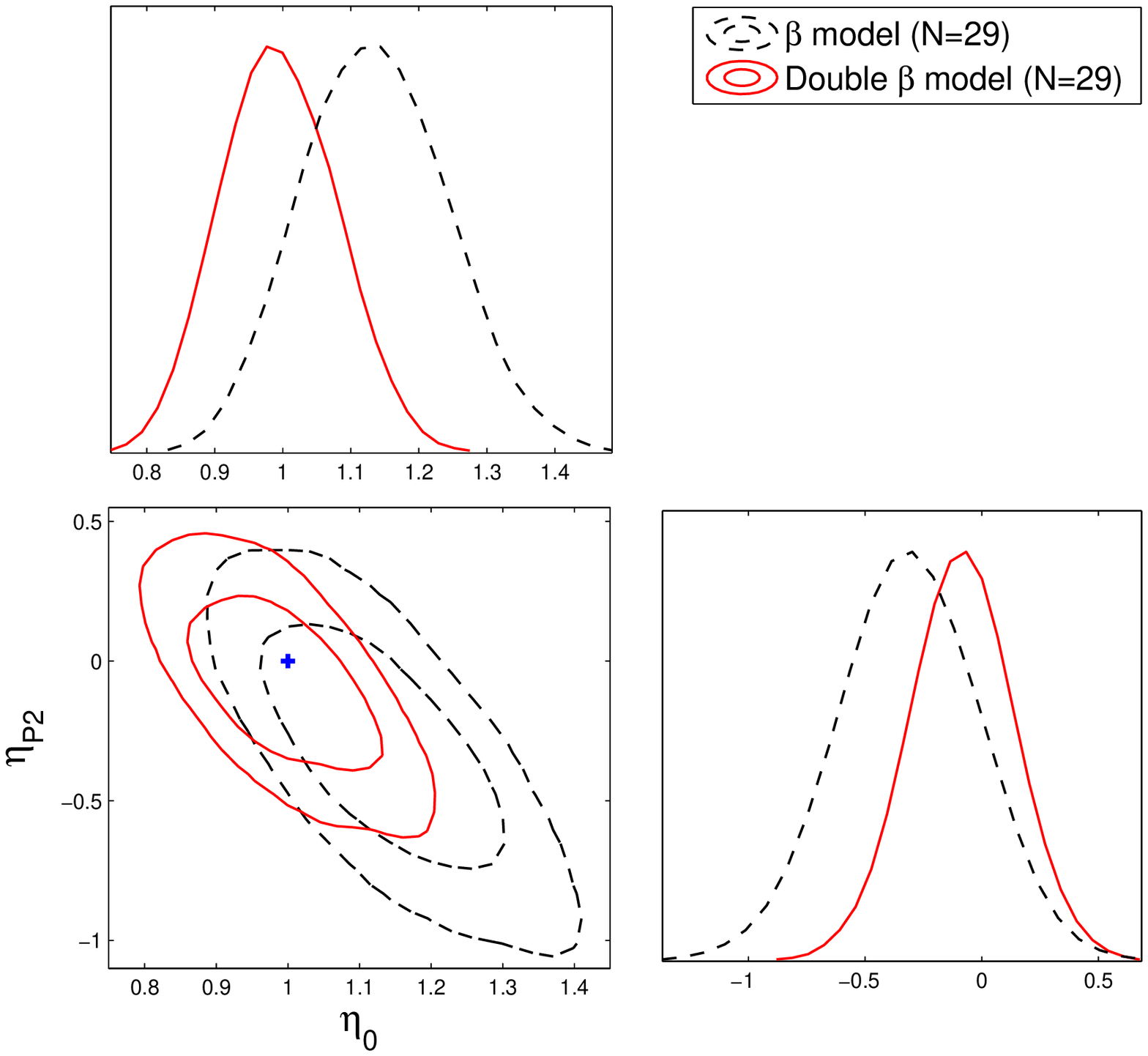}
\end{center}
\caption{ Confidence contours and marginalized likelihood
distribution functions for the parameters in
$\eta(z)=\eta_0+\eta_{P1}z$ and $\eta(z)=\eta_0+\eta_{P2}z/(1+z)$
relation. Black dashed lines and red lines correspond to the fits
obtained on the full $n=29$ sub-sample under the assumption of
isothermal $\beta$ model and non-isothermal double-$\beta$ model,
respectively. The blue cross represents the expected case when the
DDR holds exactly ($\eta_{0} = 1$, $\eta_{P1,2} = 0$).
\label{LCDM1}}
\end{figure}

\begin{figure}
\begin{center}
\includegraphics[width=1.0\hsize]{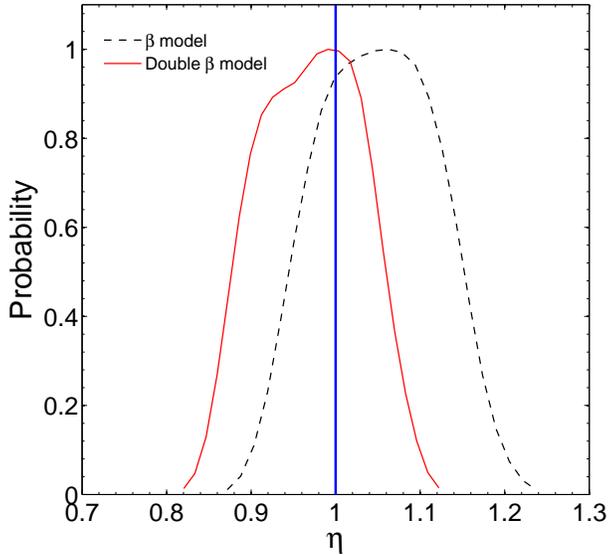}
\end{center}
\caption{Likelihood functions for $\eta$ parameter corresponding to
the isothermal $\beta$ model (black dashed line) and non-isothermal
double-$\beta$ model (red continuous line). Results are obtained
with the $n=29$ sub-sample of clusters. WMAP9 best fitted
$\Lambda$CDM parameters were taken to represent a ``true''
cosmology. The blue vertical line represents the expected case when
the DDR holds exactly. \label{1P}}
\end{figure}

\begin{figure}
\begin{center}
\includegraphics[width=1.0\hsize]{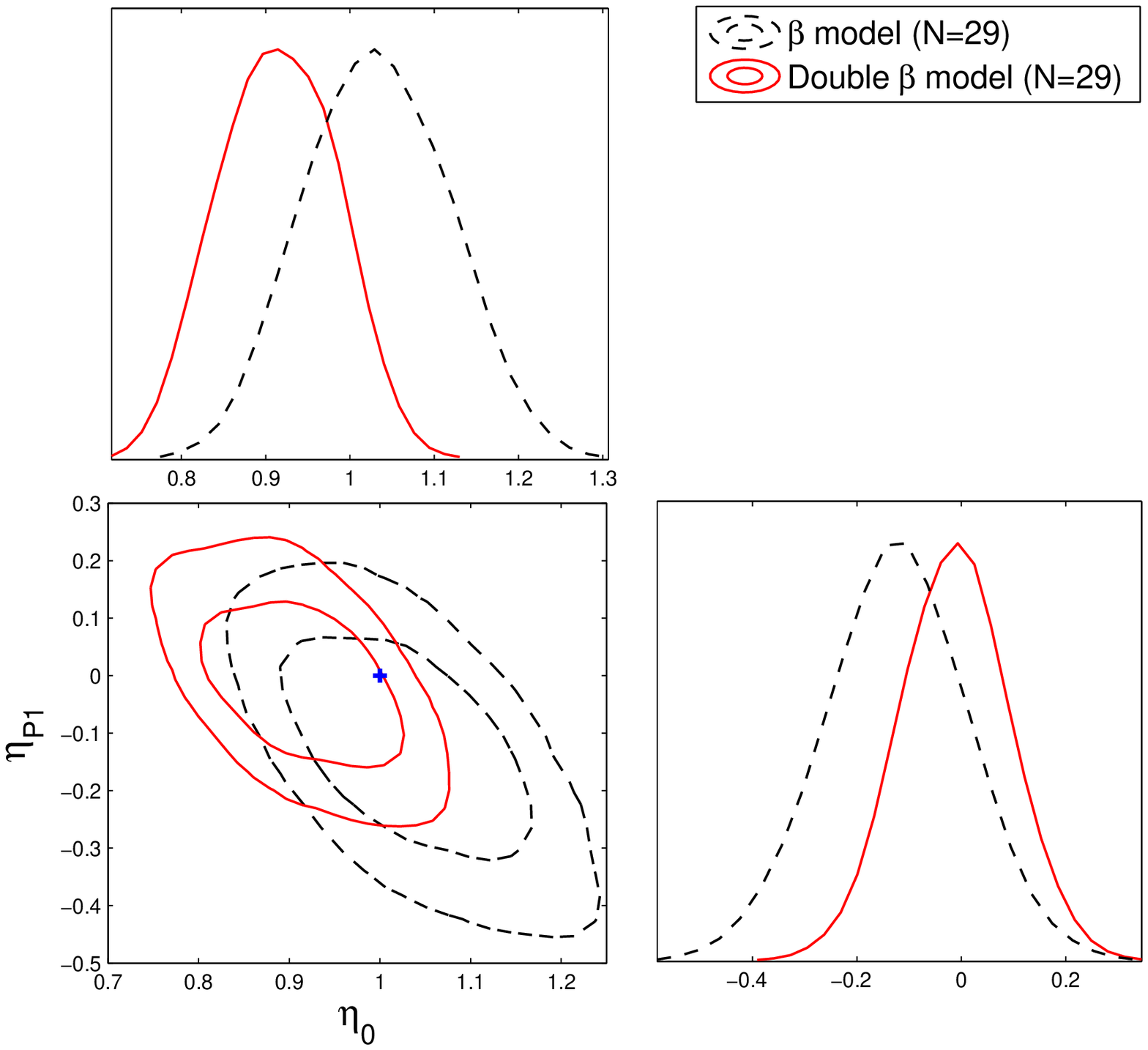}
\includegraphics[width=1.0\hsize]{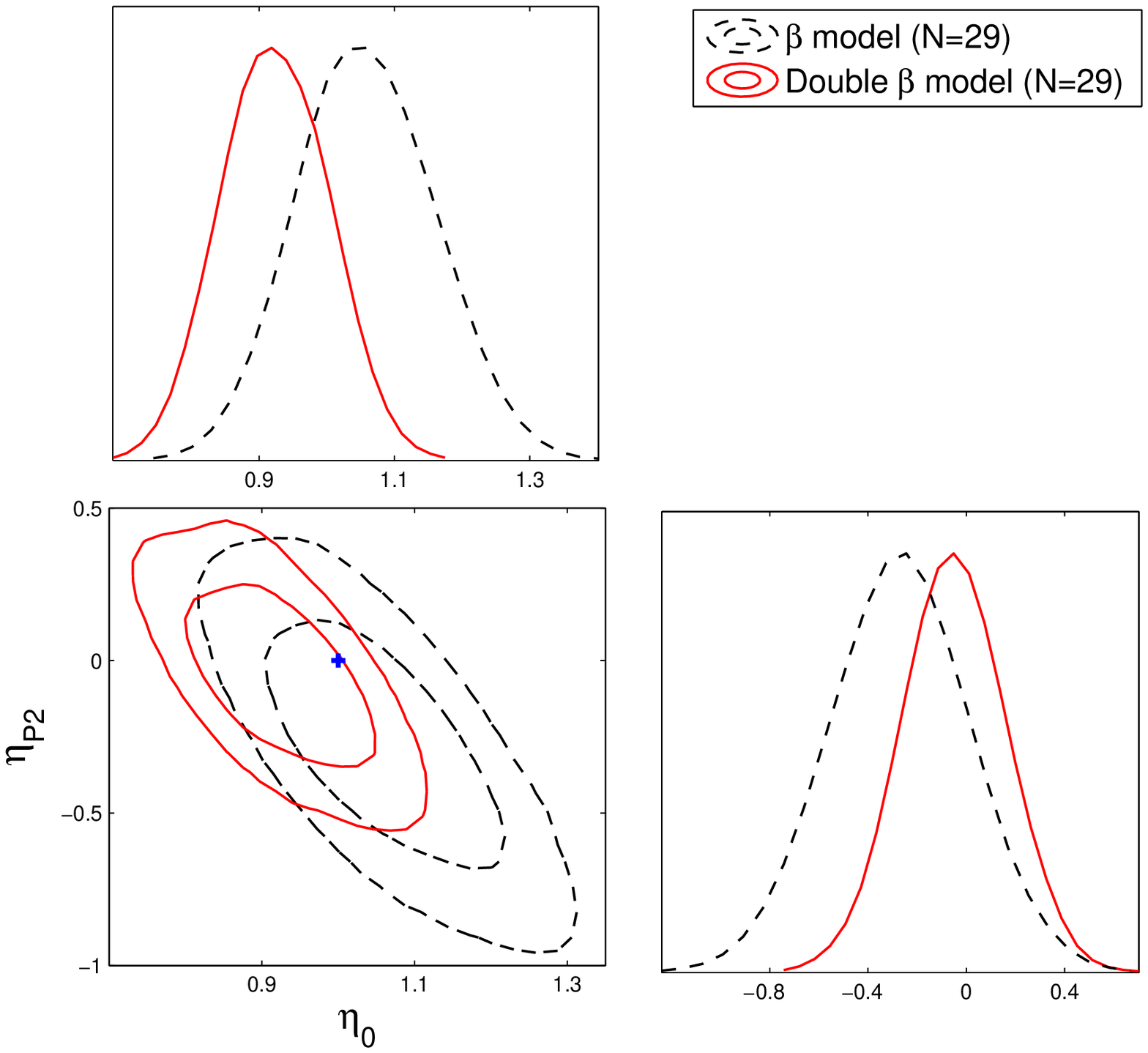}
\end{center}
\caption{ Confidence contours and marginalized likelihood
distribution functions for the $\eta$ parameters for the case when
Planck best fitted $\Lambda$CDM parameters were taken to represent a
``true'' cosmology. Black dashed lines and red lines correspond to
the fits obtained on the $n=29$ sub-sample under the assumption of
isothermal $\beta$ model and non-isothermal double-$\beta$ model.
The blue cross represents the expected case when the DDR holds
exactly. \label{LCDM2}}
\end{figure}

\begin{figure}
\begin{center}
\includegraphics[width=1.0\hsize]{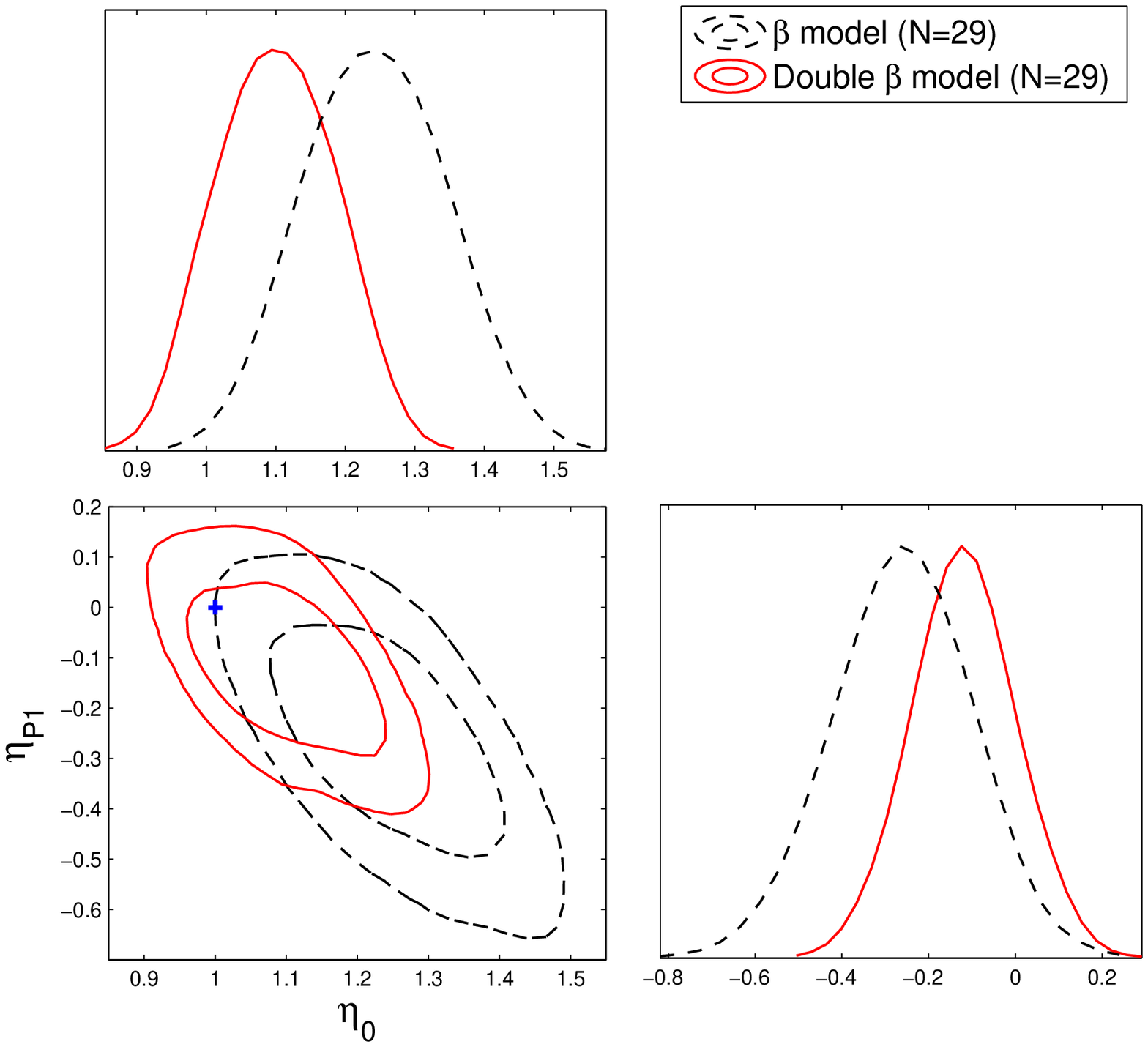}
\includegraphics[width=1.0\hsize]{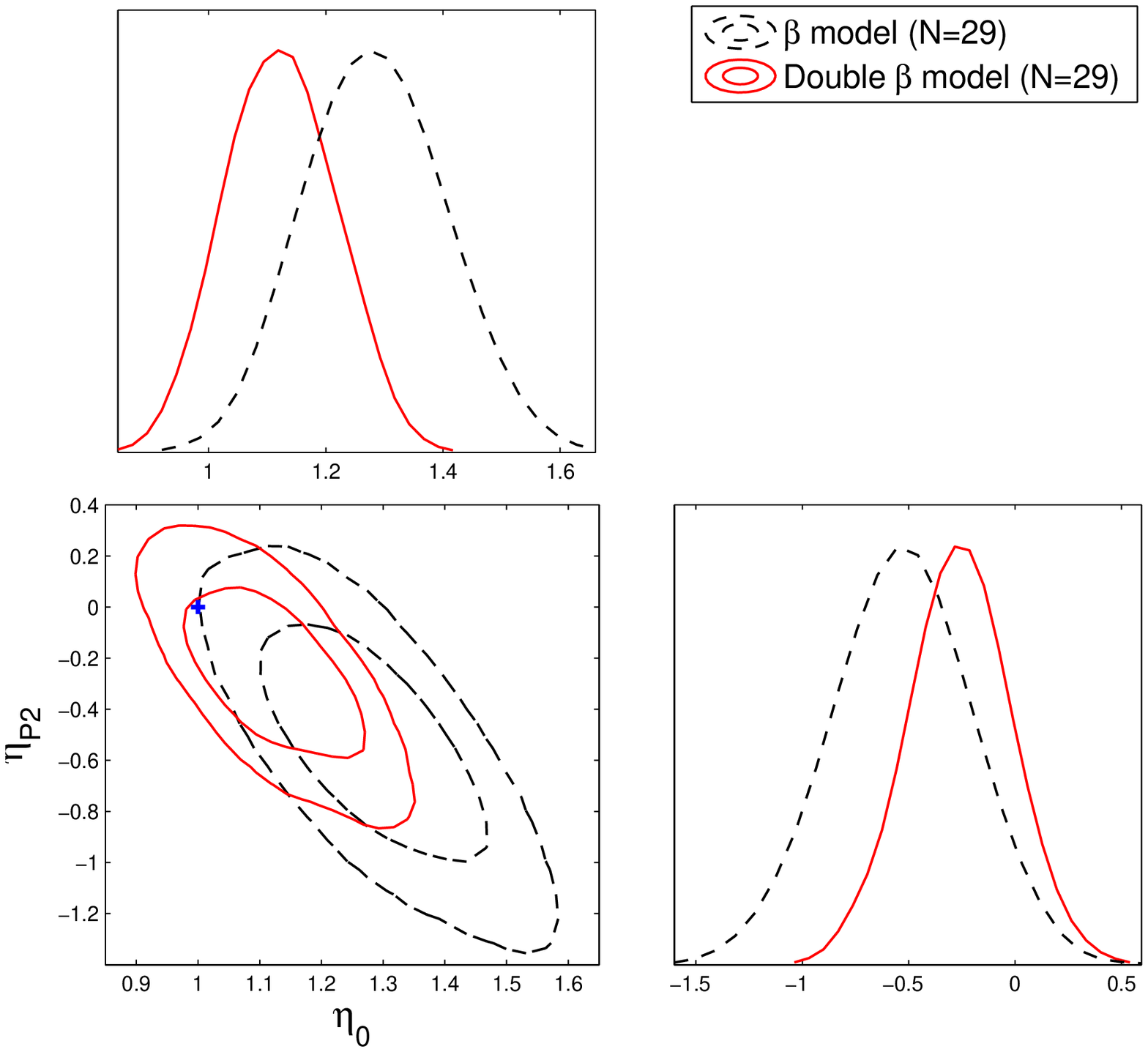}
\end{center}
\caption{ The same as Fig~\ref{LCDM2}, but for the case when XCDM
model was taken to represent a ``true'' cosmology. \label{XCDM}}
\end{figure}

According to the findings of \citet{LaRoque06}, some objects: Abell
665, ZW 3146, RX J1347.5-1145, MS 1358.4 + 6245, Abell 1835, MACS
J1423+2404, Abell 1914, Abell 2163, Abell 2204 have
 questionable reduced $\chi^2$. By excluding these objects
from the full sample we obtained a sub-sample of 29 galaxy clusters,
on which we performed a similar analysis. The results are displayed
in Fig.~\ref{LCDM1}. As compared with the previous case, one can see
that the non-isothermal double-$\beta$ model is in even much better
agreement with the DDR ($\eta_0=0.981\pm0.121$, $\eta_{P1}=-0.023\pm
0.158$; $\pm$ corresponds to $68.3\%$ CL) than the isothermal
$\beta$ model ($\eta_0=1.108\pm0.148$, $\eta_{P1}=-0.144\pm 0.387$).
When compared with the previous analysis, the incompatibility of the
isothermal $\beta$-model with the validity of DDR is clearly more
evident at $1\sigma$, although the significance of this conclusion
is not high enough from the statistical point of view.

In the case of second parametrization $\eta(z) =
\eta_0+\eta_{P2}z/(1+z)$, results for the $n=38$ full sample and
$n=29$ sub-sample, are shown in Table~\ref{tab:result}. Respective
confidence regions on the $\eta_{0}-\eta_{P2}$ plane and
marginalized likelihood distributions for the parameters are shown
in Fig.~\ref{LCDM1}. As one can see the double-$\beta$ model again
seems to be favored over the isothermal $\beta$-model.

We have also checked whether one can get tighter constraints on
$\eta$ (and hence gain more discriminative power concerning
alternative density profiles) by fixing $\eta_P=0$. The results are
shown in Fig.~\ref{1P}. One can see that the two model density
profiles are statistically compatible but again the double-$\beta$
model looks better with the mode of the likelihood coinciding with
the DDR expectation $\eta=1$.

As we already mentioned, one can contemplate other types of best
fitted ``true'' cosmology instead of the standard concordance
$\Lambda$CDM model. Therefore, we also considered the $\Lambda$CDM
model but with parameters bets fitted to the Planck data. In this
case, we only analyzed the $n=29$ sub-sample and the results are
presented on Table~\ref{tab:result} and Fig.~\ref{LCDM2}. Moreover,
since $\Lambda$CDM while useful has its own conceptual problems and
might not be the ultimate model of the Universe, we have also
considered quintessential XCDM model. In particular, we have taken
its parameters $(w,\Omega_{\rm m}=1-\Omega_{\Lambda}, H_0) = (-1.05,
0.294, 70.4 km s^{-1}Mpc^{-1})$ according to \citet{Cai14} best-fit
to Planck+WMAP9 data. The results obtained with $n=29$ sub-sample
are presented in Table~\ref{tab:result} and Fig.~\ref{XCDM}.

Even though, considering both statistical and systematic
uncertainties, we find (at the level of best fitted values) that
isothermal $\beta$-model is incompatible with the validity of DDR at
1$\sigma$, it is difficult to distinguish these two density models
by using $f_{\rm gas}$ measurements. As discussed in
\citep{LaRoque06}, the results of $f_{\rm gas}$ derived from models
including the single-$\beta$ model and double-$\beta$ model fit to
the X-ray data agree well enough to claim that the cluster core can
be accounted for by either excluding it from the fit or modeling the
gas with the double $\beta$-model.

\section{Conclusions}\label{sec:Conclusions}

Clusters of galaxies are the largest virialized objects in the
Universe. Therefore they can serve as excellent probes of cosmology:
their number density can be predicted and tested against
observations. More than that, combined X-ray and Sunyaev-Zeldovich
observations can in principle be used to measure absolute distances
to the clusters and to test cosmology (the Hubble constant, dark
energy etc.). However, for cosmological applications we need to have
at least a reliable ``proxy'' for the gas mass distribution in
clusters and this is otherwise known to be complicated (e.g. from
strong and weak lensing studies). So we need to compromise by making
assumptions like isothermal $\beta$ model or its ``offspring'' --
the non-isothermal double-$\beta$ model. In this paper we addressed
the question of which of these two proxies is more supported by the
data.

Our judgement was based on the assumed validity of the DDR --- the
distance duality relation (for which there are good reasons to
believe that it's true). To be specific, we have studied two samples
of cluster gas mass fraction data obtained from 38 X-ray luminous
galaxy clusters observed by \textit{Chandra} in the redshift range
$0.14\sim 0.89$ \citep{LaRoque06} the full sample and its $n=29$
sub-sample produced by excluding some ``suspect'' clusters.

Bearing in mind, that in practice some systematic effects might
disturb the DDR relation, we parameterized it in two ways:
$\eta(z)=\eta_{0}+\eta_{P1}z$ and
$\eta(z)=\eta_{0}+\eta_{P2}z/(1+z)$. Then we checked which of the
two ``proxy'' models for gas mass distribution (isothermal
$\beta$-model and non-isothermal double $\beta$-model) performs
better. Our result is that within standard concordance cosmology
($\Lambda$CDM1) double-$\beta$ model is marginally better respecting
the DDR (at $1\;\sigma$ level). If one takes instead $\Lambda$CDM
parameters best fitted to Planck data, both models are compatible.
However, within the quintessential XCDM cosmology double-$\beta$
model is preferred at $2\;\sigma$ level. The preference of the
double-$\beta$ model over isothermal $\beta$ model can be best seen
on marginalized distributions of $\eta_0$, $\eta_P$ parameters where
it shows up irrespectively of the cosmology assumed.

We conclude by saying that as the cluster sample size increases with
upcoming X-ray cluster surveys, we hope the method proposed in this
paper may prove useful to improve the constraints on cluster gas
mass density profiles.

\section*{Acknowledgments}
The authors are grateful to the referee for very useful comments
that allowed us to improve the paper. This work was supported by the
Ministry of Science and Technology National Basic Science Program
(Project 973) under Grants Nos. 2012CB821804 and 2014CB845806, the
Strategic Priority Research Program ``The Emergence of Cosmological
Structure" of the Chinese Academy of Sciences (No. XDB09000000), the
National Natural Science Foundation of China under Grants Nos.
11503001, 11373014 and 11073005, the Fundamental Research Funds for
the Central Universities and Scientific Research Foundation of
Beijing Normal University, China Postdoctoral Science Foundation
under grant No. 2015T80052, and the Opening Project of Key Lab of
Computational Astrophysics of Chinese Academy of Sciences. This
research was also partly supported by the Poland-China Scientific \&
Technological Cooperation Committee Project No. 35-4. M.B. obtained
approval of foreign talent introducing project in China and gained
special fund support of foreign knowledge introducing project.


\label{lastpage}

\end{document}